\documentclass[10pt, reqno]{amsart}

\usepackage{graphicx}
\usepackage{amssymb, amsfonts, verbatim}
\usepackage{epsfig, amsmath, cancel}
\usepackage[mathscr]{eucal}
 
\newtheorem{theorem}{Theorem}

\newcommand{\del}[1]{\partial_{#1}}

\newcommand{\integrals}[4]{\displaystyle\int_{#1}^{#2} #3 \, d{#4}}

\newcommand{\ga}{\alpha}
\newcommand{\gb}{\beta}

\newcommand{\gd}{\delta}
\renewcommand{\ge}{\varepsilon}
\newcommand{\gf}{\phi}

\renewcommand{\gg}{\gamma} 
\newcommand{\gh}{\eta}

\newcommand{\gl}{\lambda}
\newcommand{\gm}{\mu}
\newcommand{\gn}{\nu}

\newcommand{\gp}{\pi}
\newcommand{\gq}{\theta}
\newcommand{\gr}{\rho}

\newcommand{\gs}{\sigma}

\newcommand{\gx}{\xi}
\newcommand{\gy}{\psi}

\newcommand{\gL}{\Lambda}

\newcommand{\bbR}{\mathbb{R}}

\begin{document}

\title{Areal Foliation and AVTD Behavior in $T^2$ Symmetric Spacetimes with Positive Cosmological Constant }
\author{Adam Clausen, James Isenberg}
\address{ Department of Physics\\ 
	University of Oregon \\
	Eugene, OR, 97403, USA}
\email{aclausen@uoregon.edu}
\address{ Department of Mathematics and Institute for Theoretical Science\\ 
	University of Oregon \\
	Eugene, OR, 97403, USA}
\email{jim@newton.uoregon.edu}

\begin{abstract}
We prove a global foliation result, using areal time, for $T^2$ symmetric spacetimes with a positive cosmological constant.  We then find a class of solutions that exhibit AVTD behavior near the singularity.
\end{abstract}

\maketitle

\numberwithin{equation}{section} 

\section{Introduction}

Strong Cosmic Censorship (SCC), one of the outstanding conjectures of mathematical relativity, claims that a generic maximal globally hyperbolic solution of the Einstein equations (vacuum, or for a specified matter model) cannot be extended across a Cauchy horizon. Model studies of SCC in certain families of spacetimes with specified isometries---e.g., the spatially homogeneous solutions \cite{HomogRendall}, and the $T^3$ Gowdy spacetimes \cite{RingstromGowdy}---show that, as a step toward proving SCC in such families, it is very useful to investigate whether the solutions exhibit either local oscillatory (``mixmaster") behavior near the singularity, or the more special asymptotically velocity-term dominated (``AVTD") behavior in that region. These studies also show that the existence of a geometrically motivated coordinate system compatible with the singularity, such as the ``areal coordinates" of the $T^2$-symmetric spacetimes \cite{BCIM}, is very useful. In this paper, we show that $T^3$ Gowdy and $T^2$-symmetric solutions with a positive cosmological constant admit areal coordinates covering their globally hyperbolic regions. We also use Fuchsian methods to show that there are sets of spacetimes of this sort with AVTD behavior. 

Since the point of this work is to show that a number of the results which have been proven for $T^3$ 
Gowdy and $T^2$-symmetric vacuum solutions also hold when a positive cosmological constant is 
added, we first recall the results which hold for these two classes of vacuum spacetimes. For the $T^3$ Gowdy solutions, which are defined by the presence of a $U(1)\times U(1)$ (or $T^2$) isometry group acting on the $T^3$ spacelike foliations, and by the vanishing of the twist constants, the following has 
 been shown: (i) Global areal coordinates (for which the constant time hypersurfaces are defined to be those spacelike hypersurfaces which contain $T^2$ orbits of equal area) exist for all solutions 
\cite{Moncrief}, with the time coordinate covering the range $(0,\infty)$ except in certain special (Kasner) cases. (ii) Generically, the solutions exhibit  AVTD behavior \cite{Ringstrom}. (iii) Generic solutions have their curvature unbounded as one approaches the singularity, and cannot be extended outside the
 maximal globally hyperbolic spacetime region \cite{Ringstrom}. Hence  SCC has been proven for the $T^3$ Gowdy solutions.

For the $T^2$-symmetric vacuum solutions with non vanishing twist constants, less is known. However, it has been shown that all such (maximal, globally hyperbolic) solutions admit global areal coordinates \cite {BCIM}, again with the time covering the interval $(0,\infty)$ except for certain special (Kasner) cases \cite{t0}. It has also been shown \cite{PolarT2} that, if one restricts one's attention to polarized $T^2$-symmetric vacuum solutions, in which one of the gravitational degrees has been turned off, there is a large collection of such solutions which show AVTD behavior. It has not yet been shown that these polarized solutions generically exhibit AVTD behavior; however based on numerical and heuristic studies, it is expected that this is the case. Similar such studies \cite{Weaver} suggest that in the non polarized case, the solutions generically show mixmaster behavior. 

What happens to these results if one considers spacetimes which satisfy the Einstein equations with source fields? For the special case in which the source fields are collisionless (Vlasov) matter fields, much is known: It has been shown that both the Gowdy \cite{Andreasson} and the $T^2$ symmetric \cite{ARW} solutions with Vlasov matter admit areal coordinates, and strong cosmic censorship has been proven \cite{Miha1} in both cases. These results depend on the special features of Vlasov matter, and the techniques are quite different from those used in the vacuum case.


Motivated by  
recent observations suggesting that a large fraction of our universe consists of a component with negative pressure \cite{Spergel}, and noting that a positive cosmological constant is one possible explanation of this \cite{Peebles}, we begin in this paper to study what happens if a positive cosmological constant is present (with no Vlasov fields). We obtain results similar to those for the $T^2$ symmetric vacuum solutions. In particular, we show that $T^2$-symmetric spacetimes with a positive cosmological constant admit an areal foliation, with the time coordinate taking values in $(c,\infty)$, where $c$ is a nonnegative number.  In addition,  we find a class of $T^2$-symmetric solutions with positive cosmological constant  that exhibit AVTD behavior near the singularity.  We note that the spacetimes with AVTD behavior that we discuss here are not fully polarized, but also have  less than the maximum number of free functions. Analogous to the $U(1)$-symmetric solutions with AVTD behavior studied in \cite{IM}, they might be described as ``half polarized".  The extension of AVTD results from polarized $T^2$-symmetric solutions to half-polarized $T^2$-symmetric solutions with cosmological constant which we show here  does not in fact rely on the cosmological constant being non zero; so our results here strengthen the vacuum results of \cite{PolarT2}.

The proof of the general, global existence of  areal coordinates was first presented in \cite{BCIM} for the vacuum $T^2$-symmetric case.  In our discussion here, we follow the slightly different approach of \cite{Andreasson}, in which the global areal foliation is proved for Gowdy spacetimes with Vlasov matter.  It is reasonable that the Vlasov case would serve as the most convenient  model, since  adding a cosmological constant is equivalent to a particular choice for the stress energy tensor.  We do not here describe the complete  proof, but rather we focus on the parts of the proof which are specifically affected by the presence of the cosmological constant.    

To show AVTD behavior we follow Rendall and Kichenassamy in utilizing the Fuchsian algorithm.  An introduction to this method can be found in \cite{AnalGowdy} and \cite{PolarT2}.  In Section 4 we briefly review the algorithm and identify a set of analytic initial data whose development has AVTD behavior near the singularity.  We note that it follows from the work of Dafermos and Rendall \cite{Miha2} that $T^2$ symmetric spacetimes--vacuum, with Vlasov matter, or with positive cosmological constant--are always inextendible in the expanding direction. Hence to prove strong cosmic censorship for $T^2$ symmetric spacetimes, it is sufficient to understand the behavior of the solutions near the singularity.


\section{Field Equations}

The $T^2$-symmetric spacetimes, like the Gowdy spacetimes, admit a $T^2$ isometry with spacelike orbits; but unlike the Gowdy spacetimes, they have nonvanishing twists.  Letting $X$ and $Y$ label the Killing vector fields that generate the isometry, we find that the twists are defined as
	\begin{equation} 
		K_X=\ge_{\ga\gb\gg\gd}X^{\ga}Y^{\gb}\nabla^{\gg}X^{\gd} \quad \text{and} \quad
		K_Y=\ge_{\ga\gb\gg\gd}X^{\ga}Y^{\gb}\nabla^{\gg}Y^{\gd}.
	\end{equation}
As in \cite{BCIM} we rotate $X$ and $Y$ so that $K_X$ is zero and $K_Y=K$ is not.  This can be done without loss of generality.  The Cauchy surfaces of these spacetimes (with non vanishing twist) necessarily have $T^3$ topology\footnote{The Gowdy spacetimes, with $T^2$ isometry and vanishing twist, can have Cauchy surfaces with $T^3$, $S^3$, or $S^2 \times S^1$ topology \cite{Gowdy}. Those Gowdy spacetimes discussed here, as well as in most discussions of strong cosmic censorship in Gowdy solutions, have the $T^3$ spatial topology.},  with the coordinates $(x,y,\gq)$ all having period one.  The isometry is in the $x$ and $y$ directions where the above rotation has already been done, leaving $\gq$ as the only space direction in which the fields can vary.  Thus, the isometry reduces the problem to a $1+1$ PDE system, greatly simplifying the analysis.  

It is convenient in the global areal foliations proof to use two different coordinate systems.  Conformal coordinates are used to the past of the initial surface (towards the singularity)  and areal coordinates are used to the future of the initial surface (in the direction of expansion).  In conformal coordinates the area of the group orbits, $R$, is a function of $\gq \in S^1$ and $t\in \bbR$.  In areal coordinates we set the area function $R=t$; hence the name ``areal" coordinates.  The field equations we consider are the Einstein equations with a positive cosmological constant $\gL$,
	\begin{equation} 
		G_{\gm\gn}+\gL g_{\gm\gn} =0.
	\end{equation}
	
In \textbf{\textit{areal coordinates}} the metric takes the form
	\begin{align}\label{E:ArealMetric}
		g &= e^{2(\gh-U)}(-\ga dt^2+d\gq^2) + e^{2U}[dx +Ady+ (G+AH)d\gq ]^2  \notag \\ 
		&+ e^{-2U}t^2[dy+Hd\gq ]^2 .    
	\end{align}
Here $U, A, \ga, \gh, G,$ and $H$ are functions of $\gq\in S^1$ and $t\in \bbR^+$.  Note that the functions labeled $M_1$ and $M_2$ in \cite{BCIM} have been set to zero by an appropriate coordinate change.  The Einstein equations then are as follows\\
\\
\textbf{Constraint equations:}	
	\begin{align} 
		\frac{\gh_t}{t}&=U_t^2+\ga U_{\gq}^2+\frac{e^{4U}}{4t^2}(A_t^2+\ga A_{\gq}^2) 
		+\ga e^{2(\gh-U)}\gL +\frac{e^{2\gh}}{4t^4}\ga K^2,  \label{met1} \\ 
		\frac{\gh_{\gq}}{t}&=2U_tU_{\gq}+\frac{e^{4U}}{2t^2}A_tA_{\gq} -\frac{\ga_{\gq}}{2t\ga}, \label{met2} \\
		 \ga_t &=-4t\ga^2 e^{2(\gh-U)}\gL - \frac{e^{2\gh}}{t^3}\ga^2K^2. \label{met3}
	\end{align}
\textbf{Evolution equations:}
	\begin{align} 
		& \begin{aligned} \gh_{tt}-\ga \gh_{\gq \gq} &=\frac{\ga_{\gq}\gh_{\gq}}{2}
		+\frac{\ga_t \gh_t}{2\ga}-\frac{\ga_{\gq}^2}{4\ga}+\frac{\ga_{\gq \gq}}{2}-U_t^2
		+\ga U_{\gq}^2 \\ &+\frac{e^{4U}}{4t^2}(A_t^2-\ga A_{\gq}^2)+\ga \gL e^{2(\gh-U)} 
		- \frac{3e^{2\gh}\ga}{4t^4}K^2, \end{aligned}  \label{met4} \\ 
		&U_{tt}-\ga U_{\gq \gq}=
		-\frac{U_t}{t}+\frac{\ga_{\gq}U_{\gq}}{2}+\frac{\ga_t U_t}{2\ga} 
		+ \frac{e^{4U}}{2t^2}(A_t^2-\ga A_{\gq}^2)+ \ga\gL e^{2(\gh-U)}, \label{met5} \\ 
		& A_{tt}-\ga A_{\gq \gq}=\frac{A_t}{t}+\frac{\ga_{\gq}A_{\gq}}{2}
		+\frac{\ga_t A_t}{2\ga} -4A_tU_t +4\ga A_{\gq}U_{\gq}. \label{met6}
	\end{align}
\textbf{Auxiliary equations:}
	\begin{align}
		 & 0 = G_t +AH_t, \\ 
		 &0 = H_t-\sqrt{\ga}\frac{e^{2\gh}}{t^3}K.  
	\end{align}

In \textbf{\textit {conformal coordinates}} , where $R$ is a function of $\gq$ and $t$, we can set $\ga(\gq,t)=1$.  The metric then takes the form
	\begin{align} 
		g &= e^{2(\gh-U)}(- dt^2+d\gq^2) + e^{2U}[dx +Ady+ (G+AH)d\gq ]^2 \notag \\ 
		&+ e^{-2U}R^2[dy+Hd\gq]^2 .   
	\end{align}
The Einstein equations take the form \\
\\
\textbf{Constraint equations:}
	\begin{align} 
		&0= U_t^2+U_{\gq}^2+\frac{e^{4U}}{4R^2}(A_t^2+A_{\gq}^2)
		+\frac{R_{\gq \gq}}{R}-\frac{\gh_t R_t}{R} -\frac{\gh_{\gq} R_{\gq}}{R} 
		+\gL e^{2(\gh-U)} +\frac{e^{2\gh}}{4R^2}K^2, \\ 
		&0=2U_t U_\theta + \frac{e^{4U}}{2R^2}A_t A_\theta 
		+\frac{R_{t \theta}}{R}-\frac{\eta_t R_\theta}{R}-\frac{\eta_\theta R_t}{R} .
	\end{align}
\textbf{Evolution equations:}
	\begin{align} 
		&U_{tt}-U_{\gq \gq}=\frac{R_\gq U_\gq}{R}-\frac{R_tU_t}{R}
		+\frac{e^{4U}}{2R^2}(A_t^2-A_\gq^2) +\gL e^{2(\gh-U)}, \\ 
		&A_{tt}-A_{\gq \gq}= \frac{A_t R_t}{R}-\frac{A_\gq R_\gq}{R}+4(A_\gq U_\gq-A_tU_t), \\ 
		& R_{tt}-R_{\gq \gq}=2R \gL e^{2(\gh-U)} + \frac{e^{2\gh}}{2R^3}K^2, \\ 
		&\gh_{tt}-\gh_{\gq \gq}= U_\gq^2-U_t^2+\frac{e^{4U}}{4R^2}(A_t^2-A_\gq^2)
		+\gL e^{2(\gh-U)} - \frac{3e^{2\gh}}{4R^2}K^2.
	\end{align}
\textbf{Auxiliary equations:}
	\begin{align} 
		&0 = G_t +AH_t, \\ 
		& 0 = H_t - \frac{e^{2\gh}}{R^3}K. 
	\end{align}
We recover the equations for the vacuum Gowdy spacetimes by setting the cosmological constant and the remaining twist constant $K$ to zero, and letting both $\ga(\gq,t)=1$ and $R(\gq,t)=t$.  In this situation the constraints decouple from the evolution equations and can be effectively ignored in the course of the analysis until the end.  Adding any one of the following---nonzero twist, cosmological constant, or Vlasov matter---does not allow us to have both $\ga=1$ and $R=t$.  Hence, the constraint and evolution equations are necessarily coupled in this setting.  

Local existence for each of these coordinate systems is proven in \cite{Chrusciel}.


\section{Global Areal Foliation}

Let $\gg$ be a smooth Riemannian metric on $T^3$ and let $\gp$ be a smooth symmetric tensor on $T^3$ which together satisfy the Einstein constraint equations with positive cosmological constant $\gL$ and which are invariant under an effective $T^2$ group action. We then call $(T^3,\gL; \gg, \gp)$--which we shall sometimes abbreviate as $ (\gg, \gp)$--a set of $T^2$-symmetric Einstein-$\gL$ initial data . The following theorem, in the spirit of \cite{BCIM} and \cite
{Andreasson}, is our main result on the existence of global areal coordinates for the spacetime development of such data:


\begin{theorem}\label{T:GlobalExist}
Let $(T^3,\gL; \gg, \gp)$ be  a set of $T^2$-symmetric Einstein-$\gL$ initial data on $T^3$.  For some nonnegative constant $c$, there exists a globally hyperbolic spacetime $(M^4,g)$ such that
	\begin{enumerate}
	\item $M^4 = T^3\times (c,\infty)$
	\item $g$ satisfies the Einstein equations with positive cosmological constant $\gL$.  
	\item $M^4$ is covered by areal coordinates $(x,y,\gq,t)$ with $t\in(c,\infty)$, so the metric globally takes the form \eqref{E:ArealMetric}.
	\item $(M^4,g)$ is isometrically diffeomorphic to the maximal globally hyperbolic development of the initial data $(T^2, \gL; \gg,\gp)$.
	\end{enumerate}
	\end{theorem}

\subsubsection*{Outline of the proof} 

The basic steps of the proof follow the pattern established in \cite{BCIM} and continued in \cite{Andreasson}. In particular, the first part of the analysis involves the study of the fields to the past of the initial surface, in the contracting direction. Working in terms of conformal coordinates, we show in Section 3.1 that as long as the area function $R(\gq,t)$ stays bounded away from zero, the past maximal globally hyperbolic development in terms of conformal coordinates for this initial data, denoted $D^{-}_{\text{conf}}(\gg,\gp)$, has $t\to -\infty$.

The next part of the proof involves the establishment of a number of geometric results related to the behavior of the function $R$ in the spacetime development of the initial data. One shows that $R$ is positive everywhere in the maximal development $D(\gg,\gp)$ of the initial data, and that $R$ approaches a nonnegative value $R_0$ along any past-directed inextendible timelike path contained in $D^{-}_{\text{conf}}(\gg,\gp)$. Further, one finds that this limit $R_0$ is  the \textit{same} limit along all such paths. Thus one can identify $R_0$ with the number $c$ which appears in the statement of Theorem 1.  One now proceeds to show that  $D^{-}_{\text{conf}}(\gg,\gp)$ admits a foliation by constant $R$ Cauchy surfaces, and in addition one shows that if the conformal time $t\to -\infty$, then $D^{-}_{\text{conf}}(\gg,\gp)$ is identical to the past maximal development $D^{-}(\gg,\gp)$ of the initial data. All of these geometric results are proven in Section 5.2 of \cite{BCIM} for the vacuum case. The proofs are essentially unchanged by the addition of the cosmological constant, so we leave them out of this paper, and refer to \cite{BCIM}.


The proof in the expanding direction is more direct, as we work in areal coordinates.  We choose one of the $R=$ constant Cauchy surfaces from the past of the initial surface.  The spacetime induces data $(\gg_1,\gp_1)$ on this surface.  The future maximal globally hyperbolic development in terms of areal coordinates for this data, denoted$D^{+}_{\text{areal}}(\gg_1,\gp_1)$, has $t\to+\infty$.  This is shown here in Section 3.2.  To complete the proof, we cite another result from \cite{BCIM} which shows that $D^{+}_{\text{areal}}(\gg_1,\gp_1) \approx D^{+}(\gg_1,\gp_1)$.



\subsection{Contracting Direction}

In this section we sketch the steps of the first part of the proof, which establishes global existence in terms of conformal coordinates  in the contracting direction,  so long as the metric function $R$ stays bounded away from zero.  We follow the overall outline of the proof carried out in \cite{Andreasson}, detailing those steps in which the presence of a non zero cosmological constant $\gL$ plays a significant role. 

We recall that the \textit{local} existence of a solution in conformal coordinates has been demonstrated by Chru\'{s}ciel in \cite{Chrusciel}.  To show that the past development of the initial data extends to $t\to-\infty$ so long as $R$ stays bounded away from zero, it is sufficient  (as a consequence of Theorems 2.1 and 2.2 in \cite{Majda}) that we obtain uniform $C^2$ bounds on the metric functions.  To verify these bounds, we proceed as follows:

\subsubsection*{Step 1}  (Monotonicity of $R$ and $C^1$ bounds on $R$.)

We first note that we can use the arguments of Theorem 4.1 of \cite{Chrusciel} to show that $\nabla R$ is timelike, i.e., that $g(\nabla R,\nabla R)<0$. Those arguments, which in \cite{Chrusciel} are carried out for $T^2$-symmetric solutions with $\Lambda = 0$ and non zero twist, apply here with $\Lambda > 0$  since they rely essentially on the positivity of the twist term in the $\Lambda = 0$ version of the constraint (2.13), and not on the explicit form of this term; hence we find that the $\gL$ term in (2.13) has the same effect. An immediate consequence of this is that $R_t$ must be nonzero everywhere.  We choose $R_t>0$ so that the past does in fact coincide with the contracting direction.  

Next we define the null vectors
	\begin{equation} 
		\del{\gl}=\frac{1}{\sqrt{2}}(\del{t}-\del{\gq}); \hspace{.5in} 
		\del{\gx}=\frac{1}{\sqrt{2}}(\del{t}+\del{\gq}),
	\end{equation}
and from the evolution equation for $R$ we calculate
	\begin{equation}\label{eq1}
		R_{\gl\gx} = R \gL e^{2(\gh-U)} + \frac{e^{2\gh}}{4R^3}K^2, 
	\end{equation}
or, equivalently
	\begin{equation} 
		R_{\gx\gl} = R \gL e^{2(\gh-U)} + \frac{e^{2\gh}}{4R^3}K^2. 
	\end{equation}
Using $\gL>0$, we see that the right hand side of equation \eqref{eq1} is positive.  We can then integrate back along the null directions, $\gl$ and $\gx$, starting at $(\gq_0,t_0)$ on the initial surface, and verify that 
	\begin{align} 
		R_\gx(\gq_0+s,t_0-s) & \leq R_\gx(\gq_0,t_0), \\ 
		R_\gl(\gq_0-s,t_0-s) & \leq R_\gl(\gq_0,t_0). 
	\end{align}
Thus, for any $t<t_0$ and $\gq \in S^1$ we have  
	\begin{align} 
		R_\gx(t,\gq) & \leq \sup_\gq{R_\gx(t_0,\cdot)}, \\ 
		R_\gl(t,\gq) & \leq \sup_\gq{R_\gl(t_0,\cdot)}. 
	\end{align}
Since $R_t=1/\sqrt{2}(R_\gx+R_\gl)$, it follows that 
	\begin{equation} 
		R_t(\gq,t) < \frac{1}{\sqrt{2}}\sup_\gq(R_\gx+R_\gl)(t_0,\cdot).
	\end{equation}
Hence we have $R_t$ bounded to the past. Since  $\nabla R$ is timelike, $|R_t|>|R_\gq|$, so we have have established  $C^1$ bounds on $R$.  Noting that in deriving these results we only use the the sign of $R_{\gx\gl}$ and $R_{\gl\gx}$, it follows that the addition of a positive cosmological constant does not change the argument used in the vacuum $T^2$-symmetric case.  

\subsubsection*{Step 2}  (Bounds on $U,A,$ and their first derivatives.)

We define here the same energy-like quadratic functions $G$ and $H$ which are used in the vacuum $T^2$-symmetric case \cite{BCIM} and in the Gowdy with Vlasov matter case \cite{Andreasson}.  (There is no relation between these forms and the metric functions of the same names.)
	\begin{align} 
		G &= \frac12 R(U_t^2+U_\gq^2) +\frac{e^{4U}}{8R}(A_t^2+A_\gq^2), \\ 
		H &= RU_tU_\gq +\frac{e^{4U}}{4R}A_tA_\gq. 
	\end{align}
Taking derivatives along the null directions we find	
	\begin{align} 
		\del{\gl}(G+H) &= RU_\gx \gL e^{2(\gh-U)} - \frac12 R_\gx[U_t^2-U_\gq^2 
		+ \frac{e^{4U}}{4R^2}(-A_t^2+A_\gq^2)], \\ 
		\del{\gx}(G-H) &= RU_\gl \gL e^{2(\gh-U)} - \frac12 R_\gl[U_t^2-U_\gq^2 
		+ \frac{e^{4U}}{4R^2}(-A_t^2+A_\gq^2)]. 
	\end{align}
If we now integrate these equations along null paths, starting at $(t,\gq)$ with $t<t_0$, and ending on the $t_0$-surface, we obtain 
	\begin{align}
		[G+H](t_0,\gq-(t_0-t))-[G+H](t,\gq) &=\integrals{t}{t_0}{K_1+U_\gx T}{\gl} \label{eq2}, \\ 
		[G-H](t_0,\gq+(t_0-t))-[G-H](t,\gq) &=\integrals{t}{t_0}{K_2+U_\gl T}{\gx}  \label{eq3},
	\end{align}
where
	\begin{align} 
		K_1 &= - \frac12 R_\gx[U_t^2-U_\gq^2 + \frac{e^{4U}}{4R^2}(-A_t^2+A_\gq^2)], \\ 
		K_2 &= - \frac12 R_\gl[U_t^2-U_\gq^2 + \frac{e^{4U}}{4R^2}(-A_t^2+A_\gq^2)], \\ 
		T &= R\gL e^{2(\gh-U)}. 
	\end{align}
	
To handle the $T$ terms in  (\ref{eq2}) and (\ref{eq3}), we proceed as in the Vlasov case \cite{Andreasson}:  Recall from the evolution equation for $R$ that we have
	\begin{equation} 
		R_{\gl\gx} = R \gL e^{2(\gh-U)} + \frac{e^{2\gh}}{4R^3}K^2, 
	\end{equation}
and a similar equation for $R_{\gx\gl}$.  Integrating these along null paths we find that 
	\begin{align} 
		R_\gl (t_0,\gq+(t_0-t)) -R_\gl(t,\gq) = \integrals{t}{t_0}{[R \gL e^{2(\gh-U)} 
		+ \frac{e^{2\gh}}{4R^3}K^2  ](s,\gq+(s-t))}{s}, \\ 
		R_\gx (t_0,\gq-(t_0-t)) -R_\gx(t,\gq) = \integrals{t}{t_0}{[R \gL e^{2(\gh-U)} 
		+ \frac{e^{2\gh}}{4R^3}K^2  ](s,\gq-(s-t))}{s}. 
	\end{align}
Since we have $C^1$ bounds on $R$ (from Step 1), the left hand side of each of these equations is bounded.  Hence the right hand side is also bounded.  This implies further that the term
	\begin{equation} 
		\integrals{t}{t_0}{[T ](s)}{s} = \integrals{t}{t_0}{[R \gL e^{2(\gh-U)}](s)}{s}
	\end{equation}
is also bounded on the interval $(t,t_0]\times S^1$.

To handle the integrals of the $K_1$ and $K_2$ terms in (\ref{eq2}) and (\ref{eq3}), we proceed as in the vacuum case \cite{BCIM} by bounding them in terms of the quadratic form $G$ in the following manner.  From the definitions of $K_1$, $K_2$, and $G$ as well as from the $C^1$ bounds on $R$, we have
	\begin{equation*} 
		|K_1| \leq \frac{C G}{R} \quad \text{and} \quad  |K_2| \leq \frac{C G}{R}. 
	\end{equation*}
Then using the elementary inequalities $\pm 2ab \leq a^2+b^2$, we also have
	\begin{equation*} 
		|U_\gx| \leq (\frac{2G}{R})^{1/2}  \quad \text{and} \quad  |U_\gl| 
		\leq (\frac{2G}{R})^{1/2}. 
	\end{equation*}
We now add equations \eqref{eq2} and \eqref{eq3} and take the suprema over $\gq$.  Using various standard inequalities along with  the crude estimate $\sqrt{G}\leq1+G$, we obtain an inequality which permits the  application of Gronwall's Lemma.  This gives us a uniform bound on $\sup_\gq G(t_1,\cdot)$, provided $R$ stays bounded away from zero.  This bound on $G(t,\gq)$ gives us bounds of $|U_t|$ and $|U_\gq|$, which we can then integrate to get bounds on $U$ itself.  We then have  bounds on $|A_t|$ and $|A_\gq|$, which can themselves be integrated to obtain a uniform bound on $A$.  Hence  we have $C^1$ bounds on $U$ and $A$.  We note that the terms introduced by the presence of the (positive) cosmological constant are easily controlled by our $C^1$ bounds on $R$.  

\subsubsection*{Step 3}  ($C^1$ bounds on $\gh$.)

As in \cite{BCIM} we define the quantity
	\begin{equation} 
		\gb := \gh +\frac32\ln{(R)}.
	\end{equation}
Then using the evolution equations for $\gh$ and $R$ we derive
	\begin{equation} 
		\gb_{\gl \gx}= Z + 2\gL e^{2(\gh-U)}, 
	\end{equation}
where $Z$ contains only quantities which we have previously shown to be bounded.  Integrating this along null paths, along with the corresponding equation for $\gb_{\gx \gl}$,  we  find that 
	\begin{equation} 
		\gb_\gl(t_0,\gq+(t_0-t)) - \gb_\gl(t,\gq) 
		= \integrals{t}{t_0}{[Z+2\gL e^{2(\gh-U)}](s,\gq+(s-t)) }{s},
	\end{equation}
together with a similar equation for $\gb_\gx(t,\gq)$.  We now recall from an argument presented in Step 2  that, given $C^1$ bounds on $R$, we determine that the integral
	\begin{equation*} 
		\int R \gL e^{2(\gh-U)} \,d\gx 
	\end{equation*}
is bounded.  It follows that  $\gb_\gl$ is bounded on $(t,t_0]\times S^1$ provided $R>\ge>0$.  The same argument applies to $\gb_\gx$.  Since $\gb_t=1/\sqrt{2}(\gb_\gl+\gb_\gx)$ and $\gb_\gq=1/\sqrt{2}(\gb_\gl-\gb_\gx)$, both $\gb_t$ and $\gb_\gq$ are bounded on $(t,t_0]\times S^1$.  Given the definition of $\gb$ and the $C^1$ bounds on $R$, we see that $\gh_t$ and $\gh_\gq$ are bounded on $(t,t_0]\times S^1$.  Integration gives a bound on $\gh$ itself.  \\

We now have $C^1$ bounds on the principle metric functions.  These bounds on $U$ and $\gh$, together with the condition that  $R$ is bounded away from zero, allow us to control all terms which contain  a cosmological constant, since they always appear in the form
	\begin{equation} 
		\gL e^{2(\gh-U)} \qquad \text{ or }\qquad R\gL e^{2(\gh-U)}.
	\end{equation}
Thus the remainder of the proof in the contracting direction is unaffected by the presence of the cosmological constant and follows as in \cite{BCIM}.


\subsection{Expanding Direction}

The work we have just described shows that, for the initial surface and $T^2$-symmetric data we have chosen on it, the past (contracting direction)  of this initial surface is foliated by $R=$ constant slices. Choosing one of these slices as the new initial hypersurface (and labeling its $R=$ constant time coordinate  as $ R= t_2$ we now have induced initial data appropriate for areal coordinates, and we seek to show global existence of the evolution in terms of these coordinates in the expanding direction, to the future of $t_2$. As in the argument for the contracting direction, the key to showing global existence is to establish $C^2$ bounds for solutions of the areal form of the equations  \eqref{met1} - \eqref{met6}. We discuss how to do this in the following steps, focussing on those steps in which the presence of a positive cosmological constant plays a role.

\subsubsection*{Step1}  (Bound on $|U|$.)  

We first establish monotonicity for the energy function $E(t)$, defined by
	\begin{equation} 
		E(t) = \int_{S^1}\left[h + \sqrt{\ga}e^{2(\gh-U)}\gL + \frac{e^{2\gh}\sqrt{\ga}}{4t^4}K^2\right ] d\gq,
	\end{equation}
where
	\begin{equation} 
		h= \frac{U_t^2}{\sqrt{\ga}} + \sqrt{\ga}U_\gq^2 + \frac{e^{4U}}{4t^2}(\frac{A_t^2}{\sqrt{\ga}}
		+\sqrt{\ga}A_\gq^2).
	\end{equation}
Using the evolution equations for $U$ and $A$ and integrating by parts we determine that
	\begin{equation} 
	\frac{d}{dt}E(t)=\frac{-2}{t}\int_{S^1}\frac{U_t^2}{\sqrt{\ga}}
		+\frac{e^{4U}}{4t^2}\sqrt{\ga}A_\gq^2 
		 + \frac{e^{2\gh}\sqrt\ga K^2}{2t^4} d\gq \leq 0.
	\end{equation}
We note that if we leave the $K^2$ term out of the definition of $E(t)$,  then this new function is also  positive and monotone decreasing.  However, if we omit the $\gL$ term, then we  no longer have a monotone function, as a consequence of  the presence of $\gL$ in the evolution equation for $U$.  The situation  is analogous to the Gowdy-Vlasov case, in which the mass energy density quantity $\gr$ appears in the definition of the energy function $E$.  

We now use this behavior of $E(t)$  to establish bounds on $|\gh|$ and $|U|$.  Defining the new function $\gb$ by
	\begin{equation} 
	\gb := \gh +\frac12 \ln{(\ga)},  
	\end{equation}
we use the constraint equations for $\gh$ and $\ga$ to show that 
	\begin{equation} 
		\gb_t = t[U_t^2 + \ga U_\gq^2 + \frac{e^{4U}}{4t^2}(A_t^2 
		+ \ga A_\gq^2)] -t\ga \gL e^{2(\gh-U)} - \frac{e^{2\gh}\ga K^2}{4t^3}. 
	\end{equation}
We then immediately see that
	\begin{align} 
		\gb_t & \leq  t[U_t^2 + \ga U_\gq^2 + \frac{e^{4U}}{4t^2}(A_t^2 
		+ \ga A_\gq^2)], \label{eq4} \\ 
		\gb_t &\geq -t\ga \gL e^{2(\gh-U)} - \frac{e^{2\gh}\ga K^2}{4t^3}. \label{eq5}
	\end{align}
Clearly the presence of the positive cosmological constant does not affect the sign of the second inequality, \eqref{eq5}.  The monotonicity of $E(t)$, together with equation \eqref{eq4}, is used to get an upper bound on $\gb$.  We refer the reader to \cite{BCIM} for the details, as the argument is no different from the vacuum case. (Note, however, that $\gn =-\gb$ is used in \cite{BCIM} so the argument for an upper bound of $\gb$ here follows that for a lower bound of $\gn$ in  \cite{BCIM}.) Unlike the vacuum case, however, to verify a lower bound on $\gb$ we must first bound $|U|$, as can been seen from the $\gL$ term in \eqref{eq5}.  We accomplish this as follows: (See also \cite{Andreasson}).  From H\"{o}lder's inequality we have
	\begin{align}
		|U(t,\gq_2)-U(t,\gq_1)| &= |\integrals{\gq_1}{\gq_2}{U_\gq(t,\gq) }{\gq}| \notag \\
		&\leq \left(\integrals{\gq_1}{\gq_2}{\ga^{-1/2}}{\gq}\right)^{1/2} 	
		\left(\integrals{\gq_1}{\gq_2}{\sqrt{\ga}U_\gq^2}{\gq}\right)^{1/2}. \label{eq6}
	\end{align}
The second integral in \eqref{eq6} is clearly bounded by $(E(t_2))^{1/2}$.  To bound the first integral we use the evolution equation for $\ga$ to find
	\begin{equation} 
		\del{t}(\ga^{-1/2}) = 2t \sqrt{\ga}e^{2(\gh-U)}\gL +\frac{e^{2\gh}\sqrt\ga K^2}{2t^3}.
	\end{equation}
Integrating this over time and space we have
	\begin{align*} 
		\integrals{\gq_1}{\gq_2}{\ga^{-1/2}(t,\gq)}{\gq} 
		&= \integrals{t_2}{t}{[2s \integrals{\gq_1}{\gq_2}{\sqrt{\ga}e^{2(\gh-U)}\gL 
		+\frac{e^{2\gh}\sqrt\ga K^2}{4s^4} ]}{\gq}}{s} 
		+\integrals{\gq_1}{\gq_2}{\ga^{-1/2}(t_2,\gq)}{\gq} \\    
		&\leq \integrals{t_2}{t}{2s\tilde{E}(s)}{s} +2\pi C \\
		 &\leq \tilde{E}(t_2) (t^2-t_2^2)+2\pi C.
	\end{align*}
Thus, for all $t\in [t_2,t_3)$ and $\gq_1, \gq_2 \in S^1$, we have
	\begin{equation}\label{eq7}
		|U(t,\gq_2)-U(t,\gq_1)| \leq C(t),
	\end{equation}
for some bounded function $C(t)$ of $t$.  This controls the oscillation of $U(t,\theta)$. To control $|U(t, \theta)|$, we proceed as follows. Note that in the following,  $C(t)$ changes from one line to the next but always refers to a bounded function.  
	
We now consider
	\begin{align} 
		|\int_{S^1}U(t,\gq)\,d\gq| &=|\integrals{t_2}{t}{\int_{S^1}U_t(s,\gq)\,d\gq}{s} 
		+\int_{S^1}U(t_2,\gq)\,d\gq| \notag \\
		&\leq \integrals{t_2}{t}{\int_{S^1}|U_t(s,\gq)|\,d\gq}{s} +|C|\notag \\ 
		&\leq \integrals{t_2}{t}{\left(\int_{S^1}\sqrt{\ga}\,d\gq\right)^{1/2}\left(\int \frac{U_t^2}{\sqrt{\ga}}\,d\gq \right)^{1/2}}{s} +|C|. \label{eq8}
	\end{align}
Again, the second integral in \eqref{eq8} is bounded by $(E(t_2))^{1/2}$.  Also, as a consequence of  the constraint equation for $\ga_t$, $\ga$ is bounded and hence $\sqrt{\ga}$ is bounded.  As before, the monotonicity of $\ga$ depends on our assumption of positive $\gL$.  Thus we have the inequality
	\begin{equation}\label{eq9}
		 |\int_{S^1}U(t,\gq)\,d\gq| \leq C(t), 
	\end{equation}
for some bounded function $C(t)$.  We now use \eqref{eq7} and \eqref{eq9} to get a bound on $|U|$.  If we define $U_+:=\max_\gq U(t,\cdot)$, we find
	\begin{equation}\label{eq10}
		U_+(t)=\int_{S^1}U(t,\gq)\,d\gq +\int_{S^1}(U_+(t)-U(t,\gq))\,d\gq. 
	\end{equation}
Using \eqref{eq7}, where $\gq_2$ is the value for which $U$ achieves its maximum and $\gq_1=\gq$,  we find	\begin{equation*} 
		|U_+(t)-U(t,\gq)|\leq C(t) 
	\end{equation*}
for all $\gq\in S^1$.  Taking the absolute value of \eqref{eq10} we find
	\begin{equation} 
		| U_+(t)|\leq |\int_{S^1}U(t,\gq) \,d\gq| + \int_{S^1}C(t) \,d\gq.
	\end{equation}
Using \eqref{eq9}, we have bounded $|U_+(t)|$ on $[t_2,t_3)$.  Repeating this procedure for $U_-(t):=\min_\gq U(t,\cdot)$ we arrive at a bound on $|U_-(t)|$.  Thus, $|U|$ is uniformly bounded on $S^1\times[t_2,t_3)$.  

\subsubsection*{Step 2}  (Bounds on $\gb$, $\ga$, $\ga_t$, and $|\gh|$.)

Given the upper bound on $\gb$ and the bound on $|U|$, we see from equation \eqref{eq5} that
	\begin{equation} 
		-C(t) \leq \gb_t
	\end{equation}
for some bounded function $C(t)$.  Integration gives
	\begin{equation} 
		\integrals{t_2}{t}{-C(s)}{s} +\gb(t_2,\gq) \leq \gb(t,\gq),
	\end{equation}
and $\gb(t,\gq)$ is bounded from below.  Thus $|\gb|$ is bounded on $S^1\times[t_2,t_3)$.  

Bounds on $\ga,\ga_t,$ and $\gh$ follow almost immediately.  Recall the constraint equation for $\ga_t$, written as
	\begin{equation} 
		\del{t} \ln{(\ga)} = \frac{\ga_t}{\ga} = - 4t\gL e^{2(\gb-U)} - \frac{e^{2\gb} K^2}{t^3}.
	\end{equation}
The bounds on $|\gb|$ and $|U|$, after an integration, give us a bound on $|\ln{(\ga)}|$.  Hence, $\ga$ is bounded away from zero.  The constraint equation for $\ga_t$ immediately gives us a bound on $\ga_t$.  Finally, from the definition of $\gb$,
	\begin{equation} 
		\gb = \gh + \frac12 \ln{(\ga)},
	\end{equation}
is is clear that $|\gh|$ is also bounded on $S^1\times[t_2,t_3)$.  

The metric functions $|U|$, and $|\gh|$ are bounded on $S^1\times[t_2,t_3)$ and $\ga$ is bounded away from zero there.  In areal coordinates, all terms induced by the cosmological constant have the form
	\begin{equation} 
		\ga e^{2(\gh-U)}\gL \qquad \text{ or } \qquad 4t\ga^2 e^{2(\gh-U)}\gL.
	\end{equation}
Whenever such terms arise in the remainder of the proof they are bounded and the argument proceeds as in \cite{BCIM}.

This completes our proof of Theorem 1.


\section{Analysis in the Neighborhood of the Initial Singularity.}

One of the prime motivations for showing that a spacetime admits areal coordinates globally is so that one can use the coordinates to locate where singular behavior occurs. In particular, if one can show that a spacetime admts areal coordinates with $t$ running from zero to infinity, then one knows that there is a big bang type singularity (whatever its nature) at $t\rightarrow 0$. One can then study its properties by studying the metric, in the form  \eqref{E:ArealMetric}, in the neighborhood of $t\rightarrow 0$.

We have just proven that every $T^2$-symmetric solution with positive cosmological constant admits global areal coordinates, but have not shown that the $t$ coordinate extends to zero. Now in the vacuum $T^2$-symmetric case (with $\Lambda=0$),  it has been proven \cite{t0} that $t$ covers $(0,\infty)$ in almost all cases. The one exception is when the initial data is flat Kasner; then one may have $t>c>0$. In the $T^2$-symmetric case with Vlasov matter \cite{t02}, and in the vacuum Gowdy case \cite{Moncrief}, we also know that the areal coordinate $t$ covers all of $(0,\infty)$. While we have not yet determined whether the addition of the cosmological constant changes this, all evidence suggests that it does not, and so we shall presume in what follows that we are working with spacetimes in which  $t\rightarrow 0$.

We wish to show here that there are $T^2$-symmetric solutions with positive cosmological constant which exhibit AVTD behavior near the singular region at $t=0$. In the course of doing this, we shall in fact find that there is a much wider class of such spacetimes than expected; further this widening of the class of AVTD solutions holds for  $T^2$-symmetric solutions whether or not the cosmological constant is nonzero. Thus we extend the earlier results obtained in \cite{PolarT2}. 

To do this, we use the Fuchsian algorithm, which has in  previous work been successfully employed to show that there are AVTD solutions in the vacuum $T^3$ Gowdy family and in the polarized $T^2$-symmetric family of solutions. To understand why the Fuchsian method is effective, we recall what AVTD behavior is: Roughly speaking,  this condition holds if, near the singularity, terms involving spatial derivatives of the fields become less important in the evolution equations than those involving time derivatives.  Thus, each spacetime observer approaching the singularity effectively ``decouples" from other observers and evolves in their own ``private universe" near the singularity.  In many cases these private universes are locally Kasner spacetimes with each observer having their own Kasner exponents.

The Fuchsian algorithm is well adapted to verifying AVTD behavior in cosmological spacetimes, because it is designed for  analyzing solutions of certain types of differential equations in the neighborhood of singularities, and more specifically for showing that these solutions are bounded and well-behaved as one approaches the singularity. Hence one may proceed as follows: One first forms a new system of PDEs (labelled here the ``truncated system"), which is obtained by eliminating the spatial derivative terms from the areal form of all of the Einstein equations except for those which contain nothing but spatial derivative terms. Next, one finds functions (a parametrized set of them) which are asymptotic solutions to the truncated system, in the sense that as $t\to 0$ these functions approach solutions to the truncated system (these functions are called the ``asymptotic data"). One then writes the metric fields as sums consisting of the asymptotic data  plus remainder terms, and one substitutes these sums into the complete Einstein system. The resulting equations act as a parametrized system of PDEs for the remainder functions, with coefficients depending on the asymptotic data. If these equations can be written in the ``Fuchsian form"
	\begin{equation}\label{eq11}
		(t\del{t}+M(\gq))\vec{v} = t^\ge \vec{f}(t,\gq,\vec{v},\del{\gq}\vec{v}),
	\end{equation}
where the  vector $\vec{v}$ represents the remainder terms, where $M(\gq)$ is a matrix which depends on the spatial coordinate $\theta$ and satisfies certain positivity conditions (noted below in Step 3),  where $\ge>0$, and where $f$ satisfies certain analyticity  conditions (also noted in Step 3 below), then a solution $\vec{v}$ exists near $t\rightarrow$ zero, with $\vec{v}$ vanishing as $t$ approaches $0$. It follows (if indeed we do achieve Fuchsian form) that there are solutions of the full system of equations which approach solutions of equations in which spatial derivatives have been neglected. We thus have solutions of the Einstein equations with AVTD behavior, parametrized by the choice of the asymptotic data.  


Our experience with vacuum $T^2$-symmetric spacetimes strongly suggests that they do not generally exhibit AVTD behavior; but rather show the more complicated Mixmaster behavior (see \cite{Weaver}). The presence of a positive cosmological constant is not expected to change that. Our previous work with the vacuum case  \cite{PolarT2} suggests that AVTD behavior is to be found in polarized $T^2$-symmetric solutions, which are characterized by setting two of the initial data functions to constants. In this work, besides extending such results to the $\Lambda>0$ case, we also show that indeed the class of solutions with AVTD behavior is more general than the polarized solutions: Both with and without the cosmological constant, we obtain (using Fuchsian techniques) solutions with AVTD behavior which are ``half-polarized" in the sense that only one of the asymptotic initial data functions is set to a constant. This condition is stated more specifically below. 

To set up our results, looking toward obtaining equations in Fuchsian form, it is useful to first rewrite the areal coordinate version of the Einstein equations (2.4)-(2.9) using the the operator $D:=t\del{t}$. (We do not include equations (2.10)-(2.11), since these are essentially decoupled from (2.4)-(2.9), and thus can be handled at the end.) We obtain

	\begin{align}
		D^2U-t^2\ga U_{\gq\gq} &= \frac{1}{2\ga}(D\ga)( DU) 
		+\frac{t^2}{2}\ga_\gq U_\gq + \frac{e^{4U}}{2t^2}((DA)^2 - t^2\ga A_\gq^2)
		+t^2\ga \gL e^{2(\gh-U)}, \label{E:fuchs1}  \\ 
		D^2A- t^2\ga A_{\gq\gq} &= 2(DA) + \frac{1}{2\ga}(D\ga)( DA) 
		+\frac{t^2}{2}\ga_\gq A_\gq -4(DA)(DU) +4\ga t^2A_\gq U_\gq, \label{E:fuchs2} \\
		D\ga &= -4t^2\ga^2e^{2(\gh-U)}\gL -\frac{e^{2\gh}}{t^2}\ga^2 K^2, \label{E:fuchs3} \\ 
		D\gh &= (DU)^2 + t^2\ga U_\gq^2 + \frac{e^{4U}}{4t^2}((DA)^2 + t^2\ga A_\gq^2)
		+t^2\ga e^{2(\gh-U)}\gL +\frac{e^{2\gh}}{4t^2}\ga K^2, \label{E:fuchs4}\\ 
		\gh_\gq &= 2(DU) U_\gq +\frac{e^{4U}}{2t^2}(DA)A_\gq  -\frac{\ga_\gq}{2\ga}. \label{E:fuchs5} 
	\end{align}
	
Since equations (4.4) and (4.5) contain time derivatives (even though they are constraint equations), 
we shall treat them as evolution equations in our analysis here. We note that the vanishing of the constraint function 
	\begin{equation} 
		N\equiv \gh_\gq -2(DU) U_\gq -\frac{e^{4U}}{2t^2}(DA)A_\gq  +\frac{\ga_\gq}{2\ga}.
	\end{equation}
is preserved in time, since it follows from (4.2)-(4.5) together with the identity $\del{t}\gh_\gq - \del{\gq}\gh_t =0$ that $N$ satisfies the linear, first-order, \textit{ordinary} differential equation
	\begin{equation} 
		\del{t}N - \frac{\ga_t}{2\ga}N=0;
	\end{equation}
if $N(t_0)=0$ for any time $t_0$, then $N(t)=0$ for all time. 
	
We now carry out the AVTD argument in a series of steps:

\subsubsection*{Step 1}  (Asymptotic solutions to the truncated system.)

Eliminating all terms with  spatial derivatives in the evolution equations (4.2)-(4.5), we obtain the truncated system
	\begin{align} 
		D^2U &= \frac{1}{2\ga}(D\ga)( DU)  + \frac{e^{4U}}{2t^2}(DA)^2
		+t^2\ga \gL e^{2(\gh-U)}, \label{eq19} \\ 
		D^2A&= 2(DA) + \frac{1}{2\ga}(D\ga)( DA)  -4(DA)(DU), \label{eq20}  \\
		D\ga &= -4t^2\ga^2e^{2(\gh-U)}\gL -\frac{e^{2\gh}}{t^2}\ga^2 K^2, \label{eq21} \\ 
		D\gh &= (DU)^2  + \frac{e^{4U}}{4t^2}(DA)^2 
		+t^2\ga e^{2(\gh-U)}\gL +\frac{e^{2\gh}}{4t^2}\ga K^2. \label{eq22}
		\end{align}
We wish to show that for a class of restricted choices of the spatial functions $k(\gq), \gf(\gq), \gy(\gq)$,  $B(\gq)$, $\gl(\gq)$ and $\gg(\gq)$  (these are effectively functions of integration), the following 
\begin{align} 
		U(\gq,t) &= \frac12(1-k(\gq))\ln{t} + \gf(\gq), \label{eq23} \\ 
		A(\gq,t) &= \gy(\gq) + t^{2k(\gq)}B(\gq), \label{eq24} \\ 
		\gh(\gq,t) &= \frac14(1-k(\gq))^2\ln{t} + \gl(\gq), \label{eq25} \\ 
		\ga(\gq,t) &= \gg(\gq) \label{eq26} ,
	\end{align}
are asymptotic solutions of the truncated system (\ref{eq19})-(\ref{eq22}). First, substituting these choices into \eqref{eq19}, we have 
	\begin{equation} 
		0= 2e^{4\gf}k^2B^2t^{2k} + \gg e^{2(\gl-\gf)}\gL t^{\frac12 (k^2+3)},
	\end{equation}
which is satisfied in the limit of $t\to0$, provided $k(\gq)>0$.  We note that it is not the $\gL$ term which places this restriction on $k(\gq)$.  From equation \eqref{eq20}, we obtain
	\begin{equation} 
		4k^2t^{2k}B = 4k^2t^{2k}B,
	\end{equation}
which holds as $t\to0$ for all values of $k$.  Equation \eqref{eq21} with the functional forms \eqref{eq23}-\eqref{eq26} substituted in produces
	\begin{equation} 
		0 = -4\gL \ga^2 t^{1/2(k^2+3)}e^{2(\gl-\gf)} - \ga^2 K^2 e^{2\gl} t^{1/2(k-3)(k+1)},
	\end{equation}
which is satisfied as $t\to0$, provided $k(\gq)>3$.  Again, it is not the $\gL$ term which places the restriction on $k$, but rather the twist term $K^2$.  Finally from  \eqref{eq22} we derive
	\begin{equation} 
		\frac14(1-k)^2 = \frac14(1-k)^2 + k^2B^2e^{4\gf} t^{2k} + \gL \ga t^{1/2(k^2+3)} 
		+\frac{\ga K^2}{4} e^{2\gl} t^{1/2(k-3)(k+1)},
	\end{equation}
which is also satisfied for $k(\gq)>3$.  

We conclude that the parametrized functions \eqref{eq23}-\eqref{eq26} satisfy the truncated set (\ref{eq19})-(\ref{eq22}) for any postive $k(\gq)$ so long as the twist vanishes, and for any $k(\gq)>3$ if the twist is non zero. The value of $\Lambda$ is irrelevant. 

\subsubsection*{Step 2}  (Expansion with remainder terms.)

As noted earlier, once we have the asymptotic solutions to the truncated system, we proceed to write the solutions to the Einstein equations as the sum of these asymptotic solutions plus remainder functions. Specifically, letting $(v,w,y,\gb)$ denote these remainders, we write
	\begin{align} 
		U(\gq,t) &= \frac12(1-k(\gq))\ln{t} + \gf(\gq) + t^\ge v(\gq,t) \label{eq12}, \\ 
		A(\gq,t) &= \gy(\gq) + t^{2k(\gq)}(B(\gq) +w(\gq,t)), \label{eq13} \\ 
		\gh(\gq,t) &= \frac14(1-k(\gq))^2\ln{t} + \gl(\gq) +t^\ge y(\gq,t),  \label{eq14} \\ 
		\ga(\gq,t) &= \gg(\gq) + t^\ge \gb(\gq,t). \label{eq15}
	\end{align}
	
\subsubsection*{Step 3}  (Fuchsian system.)

We now show that if $U, A, \gh$, and $\ga$ satisfy the Einstein equations, then the remainder terms $v, w, y,$ and $\gb$ satisfy a system of the form
	\begin{equation}\label{eq16}
		(D+M(\gq))\vec{v} = t^\ge f(t,\gq,\vec{v},\del{\gq}\vec{v}).
	\end{equation}
If we can further show that, for analytic choices of $k(\gq), \gf(\gq), \gy(\gq)$,  $B(\gq)$, $\gl(\gq)$ and $\gg(\gq)$ , we have (i) the $n\times n$ matrix $M(\gq)$  satisfies the condition that, for any  constant $\gs\in(0,1)$, one has $||\gs^M||\leq C$; and (ii) the function $\vec{f}$ is continuous in time, analytic in all other arguments, and extends continuously to $t=0$, then we may apply Theorem 1 in \cite{PolarT2} to prove that there exists a unique solution that is continuous in time, analytic in space, and vanishes as $t\to0$.  This would complete the verification of AVTD behavior.

Substituting  the functions \eqref{eq12} - \eqref{eq15} into the Einstein equations  \eqref{E:fuchs1} - \eqref{E:fuchs4},  and writing the result as a first order system by setting 
	\begin{equation} 
		\vec{v} = (v_1, \ldots ,v_8) = (v,Dv, t^\ge v_\gq,w,Dw, t^\ge w_\gq,\gb,y),
	\end{equation}
we find that the PDE system satisfied by  $\vec{v}$ takes the form
	\begin{align}
		Dv_1 &= v_2, \label{eq17} \\
		\begin{split}
			Dv_2 +2\ge v_2 +\ge^2 v_1 &= t^{2-\ge}(\gg+t^\ge v_7)(-\frac12 k_{\gq\gq}\ln{t} 
			+\gf_{\gq\gq} +\del{\gq}v_3) \\
			&+ (\gg+t^\ge v_7)[2k -\ge 2t^\ge v_1 
			-2 t^{\ge}v_2)]t^{\frac12(k^2+3)-\ge}e^{2(\gl-\gf)+2t^\ge (v_8-v_1)}\gL  \\
			&-\frac{(\gg+t^\ge v_7)}{2}[\frac12(1-k)+\ge t^\ge v_1+t^\ge v_2 ] 
			t^{1/2(k-3)(k+1)-\ge}e^{2(\gl+2t^\ge v_8)}K^2\\
			&+ \frac{t^{2-\ge}}{2}(\gg_\gq + t^\ge \del{\gq}v_7)(-\frac12 k_\gq\ln{t} + \gf_\gq +v_3) \\
			&+ \frac{t^{2k-\ge}}{2}e^{4\gf+4t^\ge v_1} (2k (B+v_4) + v_5)^2  \\
			&- \frac{t^{2-2k-\ge}}{2}e^{4\gf+4t^\ge v_1} (\gg+t^\ge v_7)(\gy_\gq 
			+ t^{2k}2k_\gq \ln{t}(B+v_4) +t^{2k}(B_\gq+w_\gq))^2, 
		\end{split} \\
		Dv_3 &= t^\ge \del{\gq}(\ge v_1+v_2), \\
		Dv_4 &= v_5, \\
		\begin{split}
			Dv_5 + 2k v_5 &= t^{2-2k}(\gg+t^\ge v_7)[\gy_{\gq\gq} +
			t^{2k}(B+v_4)((2k_\gq\ln{t})^2+2k_{\gq\gq}\ln{t}) \\
			&+ 4t^{2k}k_\gq\ln{t}(B_\gq+\del{\gq}v_4)] + t^2(B_{\gq\gq}+w_{\gq\gq})(\gg+t^\ge v_7) \\
			&- 2 (\gg+t^\ge v_7)(2k(B+v_4)+v_5 ) t^{\frac12(k^2+3)}e^{2(\gl-\gf)
			+2t^\ge (v_8-v_1)}\gL \\
			&- \frac{(\gg+t^\ge v_7)}{2} t^{1/2(k-3)(k+1)}e^{2(\gl+2t^\ge v_8)}[2k(B+v_4)+v_5 ] K^2\\
			&+ \frac{t^{2-2k}}{2}(\gg_\gq +t^\ge \del{\gq} v_7) [\gy_\gq +t^{2k}2k_\gq\ln{t}(B+v_4) 
			+t^{2k}(B_\gq+ \del{\gq}v_4)] \\
			&+4t^{2-2k}(\gg+t^\ge v_7)(\gy_{\gq} + t^{2k}2k_\gq \ln{t}(B+v_4) 
				+ t^{2k}(B_\gq+\del{\gq}v_4)) (-\frac12 k_\gq \ln{t} +\gf_\gq + v_3)\\
			& -8kt^\ge(B+v_4)(\ge v_1+v_2) -4t^\ge v_5(\ge v_1+v_5),
		\end{split} \\
		Dv_6 &=  t^\ge \del{\gq}(\ge v_4+v_5), \\
		(D+\ge)v_7 &= -4\gL(\gg +t^\ge v_7)^2 t^{\frac12(k^2+3)-\ge}e^{2(\gl-\gf)+2t^\ge (v_8-v_1)} \notag \\
		&-(\gg+t^\ge v_7)^2 K^2  t^{1/2(k-3)(k+1)-\ge}e^{2\gl+2t^\ge v_8}, \\
		\begin{split}\label{eq18}
			(D+\ge)v_8 &= t^\ge(\ge v_1 +v_2)^2 + \ge(1-k)v_1 + (1-k)v_2 \\
			&+  t^{2-\ge}(\gg+t^\ge v_7)(-\frac12 k_\gq \ln{t} +\gf_\gq + v_3)^2 \\
			&+ \frac{t^{2k-\ge}}{4}e^{4\gf+4t^\ge v_1}(2k (B+v_4) +v_5)^2 \\
			&+ \frac{t^{2-2k-\ge}}{4}e^{4\gf+4t^\ge v_1}(\gg+t^\ge v_7)(\gy_{\gq} 
			+ t^{2k}2k_\gq \ln{t}(B+v_4) + t^{2k}(B_\gq+\del{\gq}v_4))^2 \\ 
			&+ \gL (\gg+t^\ge v_7)t^{\frac12(k^2+3)-\ge}e^{2(\gl-\gf)+2t^\ge (v_8-v_1)}\\
			&+ \frac{K^2(\gg +t^\ge v_7)}{4} t^{1/2(k-3)(k+1)-\ge}e^{2\gl+2t^\ge v_8}.
		\end{split}
	\end{align}

As written, the system \eqref{eq17} - \eqref{eq18} is of the Fuchsian form \eqref{eq16}. However  the right hand side does not satisfy the necessary conditions on $\vec{f}$; that is, $\vec{f}$ does not extend continuously to $t=0$.  The equations for $v_2,v_5$, and $v_8$ have terms that would require $k<1$.  As stated earlier, terms involving the twist constant $K$ require $k(\gq)>3$.  All the terms requiring $k<1$ involve either $\gy_\gq$ or $\gy_{\gq\gq}$, and can thus  be eliminated by choosing $\gy(\gq)=$ constant in \eqref{eq13}.  Thus if we assume $\gy=$ constant and $k(\gq)>3$, then $\vec{f}$ extends continuously to $t=0$.  The function $\vec{f}$ is also seen to be analytic in the other arguments, provided the functions $k(\gq),\gf(\gq)$, and $B(\gq)$ are analytic.  The matrix $M(\gq)$ is
	\begin{equation} 
			M= 
		\begin{pmatrix}
			0 & -1 & 0 & 0 & 0 & 0 & 0 & 0 \\
			\ge^2 & 2\ge & 0 & 0 & 0 & 0 & 0 & 0 \\
			 0 & 0 & 0 & 0 & 0 & 0 & 0 & 0 \\
			 0 & 0 & 0 & 0 & -1 & 0 & 0 & 0 \\
			 0 & 0 & 0 & 0 & 2k & 0 & 0 & 0 \\
			  0 & 0 & 0 & 0 & 0 & 0 & 0 & 0 \\
			 0 & 0 & 0 & 0 & 0 & 0 & \ge & 0 \\
			 \ge(k-1)& (k-1)& 0 & 0 & 0 & 0 & 0 &\ge
		\end{pmatrix}.
	\end{equation}
That $M(\gq)$ satisfies the bound $||\gs^M||\leq C$ for $0<\gs<1$, can been seen by direct computation. 

The argument just completed shows that for every choice of the analytic functions $k(\gq), \gf(\gq), \gy(\gq)$,  $B(\gq)$, $\gl(\gq)$ and $\gg(\gq)$  satisfying the conditions $k(\theta)>3$ and $\gy(\gq)=$ constant, we have a $T^2$-symmetric solution of the Einstein equations, either with $\Lambda>0$ or with $\Lambda=0$, which has AVTD behavior. Comparing with the earlier work \cite{PolarT2} on vacuum polarized $T^2$-symmetric, we see that we have a wider class of solutions with AVTD behavior, since here we allow $B(\gq)$ to be any analytic function, while in \cite{PolarT2}, we set it to zero. This result is consistent with the expectation that generic $T^2$-symmetric spacetimes do not have AVTD behavior, but rather have the more complicated Mixmaster behavior.


\section*{Acknowledgements}
A.C. thanks Beverly Berger for useful Mathematica files and Paul Allen for helpful comments on the draft.
This work is partially supported by a generous grant from Les Matson and NSF grant PHY-0354659 at the University of Oregon.


\end{document}